\documentstyle[rotating,english,preprint,aps,amssymb,graphicx]{revtex}

\newcommand{\be}{\begin{equation}}
\newcommand{\ee}{\end{equation}}
\newcommand{\bea}{\begin{eqnarray}}
\newcommand{\eea}{\end{eqnarray}}
\newcommand{\ruo}{\rule{0pt}{19pt}}
\newcommand{\ruu}{\rule[-10pt]{0pt}{19pt}}

\begin{document}
\draft

\title{Theory for the reduction of products of spin operators}

\author{P.J. Jensen$^*$ and F. Aguilera-Granja}

\address{Instituto de F\'\i sica ``Manuel Sandoval Vallarta'', \\
Universidad Aut\'onoma de San Luis Potos\'\i\\
San Luis Potos\'\i, 78000 S.L.P., Mexico}

\date{\today}
\maketitle

\begin{abstract} 
In this study we show that the sum of the powers of arbitrary 
products of quantum spin operators such as 
$(S^+)^l(S^-)^m(S^z)^n$ can be reduced by one unit, if 
this sum is equal to $2S+1$, $S$ being the spin quantum number.  
We emphasize that by a repeated application of this 
procedure \em all \em arbitrary spin operator products 
with a sum of powers larger than $2S$ can be replaced by a combination
of spin operators with a maximum sum of powers not larger than 
$2S$. This transformation is exact. All spin operators must belong 
to the same lattice site. By use of this procedure the 
consideration of single-ion anisotropies and the investigation of 
the magnetic reorientation within a Green's function theory are 
facilitated. Furthermore, it may be useful for the study of 
time dependent magnetic properties within the ultrashort (fsec) time 
domain.\\  
PACS: 05.50.+q, 75.10.Jm, 75.30.Gw \\  
\vspace{1cm}
\end{abstract} 

A renewed interest in the properties of spin operators and their 
products is caused by the investigation of the magnetic anisotropy 
and other magnetic properties of thin films and multilayers \cite{BlH94}. 
In general the lattice anisotropies are defined by the expansion of 
the free energy $F(T,\Theta,\Phi)$ in powers of the direction cosines, 
or by a complete set of spherical harmonics Y$_{lm}(\Theta,\Phi)$ 
\cite{MSK98}, with $\Theta$ and $\Phi$ being the polar and the 
azimuthal angles, respectively. The coefficents of this expansion 
are temperature dependent, these \em effective \em anisotropy 
coefficients are measurable quantities \cite{LRP93,FMA97}. To compare 
the measurements with first principle electronic calculations 
of the \em microscopic \em anisotropy parameters at $T=0$ 
\cite{HBW97,UZW99}, the temperature dependences of the effective 
anisotropies have to be known \cite{Cal66,JeB98,foot}. 

The magnetic lattice anisotropy is mainly a single-ion property. 
Corresponding to the atomic lattice symmetry, they can be represented 
in the Hamiltonian by powers of the components of the 
three-dimensional Heisenberg spin $S^\alpha$, $\alpha=x,y,z$, and 
mixed products of these spin operators. By use of this generalized 
Heisenberg-type Hamiltonian the magnetic properties, in particular 
the effective anisotropies, can be determined by the usual rules of 
Statistical Mechanics \cite{Cal66,JeB98,Mil95,HuU97}. Note that in 
addition to the \em diagonal \em operators such as $(S^z)^n$ or 
$(S^-S^+)^n$, also \em non-diagonal \em operators may occur. 
Here we have introduced the non-hermitean raising and lowering spin 
operators $S^\pm=S^x\pm i\,S^y$, assuming the quantization axis along
the $z$- direction. This direction may be the easy axis of the 
magnetization determined by the anisotropies, for example.  
Due to an external magnetic field the 
spins are tilted away from the easy axis. Then by application of the 
usual rotation properties of spin operators \cite{BCS72}, non-diagonal 
spin operators are introduced in the Hamiltonian. 

For a single-spin problem, the respective expectation values of 
spin operators can be easily deduced from simple diagonalization 
\cite{JeB98,Mil95,HuU97}. However, in the case of interacting extended
systems, in which the magnetic moments are coupled by the exchange 
interaction, approximations are necessary. The exchange coupling is 
in general much stronger than the involved anisotropies. Within 
a mean field approximation the system is reduced to an effective 
single-particle system, subject to a molecular field due to the 
exchange coupling \cite{JeB98,Mil95,HuU97}. This approach yields a 
bad discription of the magnetization in particular for low-dimensional 
systems, since collective excitations (spin-waves) of the Heisenberg 
magnet are neglected. These spin fluctuations are considered by 
spin-wave theories. For instance, the Holstein-Primakoff 
approach has been applied frequently \cite{HoP40,ErM91}, which 
is valid for low temperatures. On the other hand, the Green's 
function method \cite{Cal63,Tya67,FJK99} and the Schwinger-Boson 
approximations \cite{TGH98,TiJ00} can be used also at higher 
temperatures, since they consider partly interactions between 
spin-waves. The Green's function method is a 
hierachical approach, i.e.\ in the respective equation of motion 
appear higher order Green's functions consisting of a 
larger number of spin operators \cite{Cal63,Tya67}. In order to 
obtain a closed set of equations for the Green's functions, the higher
order ones have to be decoupled, i.e.\ approximated by lower order 
ones at a certain level of hierarchy. 

The single-ion lattice anisotropies are taken into account by \em 
local \em Green's functions, i.e.\ consisting only spin operators of 
a single lattice site. Such local Green's functions with non-diagonal 
operators appear within the spin wave treatment of the magnetic 
reorientation \cite{FJK99}, or in the case of a noncollinear magnetism. 
In this study 
we will show that under certain conditions the higher order local  
Green's functions appearing in the equation of motion, or the 
respective expectation values as obtained from the spectral theorem, 
can be replaced by lower order ones. This will be the case for general
products of spin operators such as $(S^x)^l(S^y)^m(S^z)^n$ or 
$(S^+)^l(S^-)^m(S^z)^n$, if the sum of their powers (the 'total power') 
is $l+m+n=2S+1$, $S$ being the (integer or half-integer) spin quantum 
number. In this case the total power can be reduced by one unit, 
resulting then in lower order Green's functions or expectation values. 
Any combination of spin operators can be reduced, in particular also 
non-diagonal ones such as $(S^\pm)^m(S^z)^{2S+1-m}$. 
This procedure is exact. The reason for 
this property is the fixed magnitude of the spin operator {\bf S}, 
i.e.\ $({\bf S})^2=(S^x)^2+(S^y)^2+(S^z)^2=S(S+1)$, thus the three 
components are not independent of each other. 
Furthermore, by a repeated application of this procedure also 
spin operator products with a sum of powers larger than $2S+1$ can be 
reduced correspondingly to a total power equal to $2S$. We are not 
aware in the literature about such a reduction in general.  
The outlined procedure seems to be a special case of the relationships
between irreducible unit tensor operators with different rank
\cite{BCS72,BMO86}. In a forthcoming paper we will investigate this
connection into greater detail. 

Quantum mechanical spins are characterized by the commutation relations 
$[S^+,S^-]=2S^z$ and $[S^z,S^\pm]=\pm\,S^\pm$, in units of 
$\hbar\equiv1$. As mentioned, the spin operators under consideration 
belong to the same Heisenberg spin {\bf S}, thus a particular lattice 
site index is omitted here. In case of time dependent spin operators 
the same time for all operators is assumed. The description of 
the reduction of the spin operator products 
is performed within several steps. As an example we show the 
respective procedure of the particular product $(S^-)^m(S^z)^{2S+1-m}$. 
Other combinations can be obtained similarly, or by application 
of the commutation relations between spin operators. 

\noindent 1) First we state the general property for the raising 
and lowering operators, 
\be (S^-)^{2S+1}=(S^+)^{2S+1}=0\,. \label{e1} \ee

\noindent 2) From the the general identity \cite{Cal63,Tya67}
\be \prod_{i=-S}^S\;(S^z-i) =0\,, \label{e2} \ee
one can deduce the first reduction in the total power of spin 
operator products, namely 
\be (S^z)^{2S+1}=\sum_{i=0}^{2S}\,\alpha_i^{(S)}(S^z)^i\,, 
\label{e3} \ee
the $\alpha_i^{(S)}$ being rational coefficients dependent on $S$, 
which can be determined for each $S$ with the help of a recursion 
relation from coefficients of lower $S$. In particular one obtains  
\bea {\rm integer}\;\;S: &&
(S^z)^{2S+1}=S^z\,\sum_{i=0}^{S-1}\,\alpha_{2i}^{(S)}\,(S^z)^{2i}\,, 
\nonumber \\ 
{\rm half-integer}\;\;S: &\hspace{1cm}&  
(S^z)^{2S+1}=\sum_{i=0}^{S-1/2}\,\alpha_{2i}^{(S)}\,(S^z)^{2i}\,. 
\label{e4} \eea
The coefficients $\alpha_i^{(S)}$ for the minimal and the maximal 
index $i$ can be given in closed form,  
\bea {\rm integer}\;\;S: & &
\alpha_0^S=-S^2\,\alpha_0^{(S-1)}=
(-1)^{S-1}\prod_{i=0}^{S-1}\;(S-i)^2\,, \nonumber \\ 
&&\alpha_{2S-2}^S=\alpha_{2S-4}^{(S-1)}+S^2=\sum_{i=0}^S\;i^2\,; 
\nonumber \\ {\rm half-integer}\;\;S: & \hspace{1cm} & 
\alpha_0^S=-S^2\,\alpha_0^{(S-1)}=
(-1)^{S-1/2}\prod_{i=0}^{S-1/2}\;(S-i)^2\,, \nonumber \\ 
&&\alpha_{2S-1}^S=\alpha_{2S-3}^{(S-1)}+S^2=
\sum_{i=0}^{S-1/2}\;(i+\frac{1}{2})^2 \,. \label{e6} \eea
For the other indices the recursion relations are given by 
\be \alpha_{2i}^S=\alpha_{2i-2}^{S-1}-S^2\,\alpha_{2i}^{S-1}\;, 
\label{e5} \ee
with $2i=2,\ldots,2S-4$ for integer $S$, and $2i=2,\ldots,2S-3$ for 
half-integer $S$. 

\noindent 3) Consider now the general spin identity
\be (S^z)^2=S(S+1)-S^z-S^-S^+\,. \label{e7} \ee  
For the determination of the products $(S^-)^m(S^z)^{2S+1-m}$ multiply 
Eq.(\ref{e7}) \em from the right \em subsequently by 
$(S^z)^l$, $l=2S-1,\ldots,1$. Calculate at first for $l=2S-1$ with the 
help of Eq.(\ref{e3}) the product 
\be (S^z)^{2S}=\sum_{i=0}^{2S-1}\,\beta_i^{(S,2S-1)}\,(S^z)^i
-S^-S^+\,(S^z)^{2S-1}\,, \label{e8} \ee
the coefficients $\beta_i^{(S,2S-1)}$ depend on the $\alpha_i^{(S)}$ 
and have to be determined for each $S$. Calculate then subsequently 
for $l=2S-2,2S-3,\ldots,1$ the products
\be (S^z)^{l+1}=\sum_{i=0}^{l}\,\beta_i^{(S,l)}(S^z)^i
+S^-S^+\sum_{i=l}^{2S-1}\,\gamma_i^{(S,l)}(S^z)^i\,, \label{e9} \ee
with corresponding coefficients $\beta_i^{(S,l)}$ and 
$\gamma_i^{(S,l)}$ dependent on $S$, with 
$\gamma_{2S-1}^{(S,2S-1)}=-1$. Within this procedure the product 
$(S^z)^{l+2}$ appearing for each value of $l$ has to be replaced by 
the result of the preceeding calculation for $l+1$. Finally, for 
$l=0$ one obtains an expression for $S^z$ given by 
\be S^z=\beta_0^{(S,0)}
+S^-S^+\sum_{i=0}^{2S-1}\,\gamma_i^{(S,0)}\,(S^z)^i\,. \label{e10} \ee
Eqs.(\ref{e3}), (\ref{e8}), (\ref{e9}), and (\ref{e10}) form a set 
of equations for the powers of $S^z$ ranging from 0 to $2S+1$. 
The products $S^-S^+(S^z)^i$ remain unchanged at present. 
These equations can be viewed to represent relations between spin 
operators as well as between the respective expectation values. 

\noindent 4) By application of Eqs.(\ref{e8}) -- (\ref{e10}) we 
are now in the situation to reduce the sum of powers of the 
spin operator products 
$(S^-)^m(S^z)^{2S+1-m}$ by one unit. Consider at first the product 
$(S^-)^{2S}S^z$ ($m=2S$), and replace the operator $S^z$ by 
Eq.(\ref{e10}). Due to $(S^-)^{2S+1}=0$, Eq.(\ref{e1}), the terms 
consisting $S^-S^+$ on the right hand side of Eq.(\ref{e10}) vanish, 
yielding \be (S^-)^{2S}S^z=\delta_0^{(S,2S)}\,(S^-)^{2S}=
\beta_0^{(S,0)}\,(S^-)^{2S}=S\,(S^-)^{2S}\,. \label{e11} \ee
The latter identity is obtained from inspection. The total power 
of the product $(S^-)^{2S}S^z$ has thus reduced by one unit. 

\noindent 5) Consider now subsequently the products 
$(S^-)^m(S^z)^{2S+1-m}$ with $m=2S-1,\ldots,1$. As in the preceeding 
step, replace the operator $(S^z)^{2S+1-m}$ with the help of 
Eq.(\ref{e9}) for $l+1=2S+1-m$. Use then the result for the preceeding
spin operator product $(S^-)^{m'}(S^z)^{2S+1-m'}$ with $m'=m+1$. 
Within the course of these calculations one has to commute 
the product $S^+(S^z)^i$ by use of $S^+(S^z)^i=(S^z-1)^iS^+$, 
replace the product $S^-S^+$ by the identity Eq.(\ref{e7}), and  
finally solve the result for $(S^-)^m(S^z)^{2S+1-m}$. Together with 
Eqs.(\ref{e3}),(\ref{e11}) a set of equations is obtained for the products 
\be (S^-)^m(S^z)^{2S+1-m}=(S^-)^m\sum_{i=0}^{2S-m}\,\delta_i^{(S,m)}\,
(S^z)^i\,, \label{e12} \ee
for $0\le m\le2S$. As can be seen, the products of the left hand 
side of Eq.(\ref{e12}), with the total power equal to $2S+1$, 
have been replaced by a sum of spin operator products whose maximal 
sum of powers is $2S$. Thus the total powers of the original 
spin operator products have been reduced by one unit. The rational 
coefficients $\delta_i^{(S,m)}$ depend on the spin quantum number $S$. 
For $m=0$ they are given by the coefficients $\alpha_i^{(2S)}$
of Eq.(\ref{e3}). In Table 1 we present results for $\delta_i^{(S,m)}$ 
for several values of the spin quantum number $S$. For the maximal 
index $i=2S-m$ one obtains from inspection 
$\delta_{2S-m}^{(S,m)}=m(2S+1-m)/2$, furthermore   
for integer $S$ one has $\delta_0^{(S,1)}=0$. As an example for the
reduction procedure and for the use of the
Table 1 we show explicitely the respective results for the case $S=2$. 
The products of spin operators $(S^-)^m(S^z)^{2S+1-m}$ with a total
power of $2S+1=5$ is replaced by a series with increasing
powers of $S^z$, and a maximum total power of $2S=4$. 
\bea (S^z)^5&=&-4\;S^z+5\;(S^z)^3 \nonumber \\ 
S^-\,(S^z)^4&=&-2\;S^-\,S^z+S^-\,(S^z)^2+2\;S^-(S^z)^3 \nonumber \\
(S^-)^2\,(S^z)^3&=&-2\;(S^-)^2\,S^z+3\;(S^-)^2\,(S^z)^2 \nonumber \\
(S^-)^3\,(S^z)^2&=&-2\;(S^-)^3+3\;(S^-)^3\,S^z \nonumber \\
(S^-)^4\,S^z&=&2\;(S^-)^4 \nonumber \\
(S^-)^5&=&0 \nonumber \eea

\noindent 6) To consider the spin operator products 
$(S^z)^{2S+1-m}(S^+)^m$, Eq.(\ref{e7}) has to be multiplied \em from 
the left \em by $(S^z)^l$. Eqs.(\ref{e8}) -- (\ref{e10}) remain 
unchanged if $S^-S^+(S^z)^i$ is substituted by $(S^z)^iS^-S^+$, since 
$S^-S^+$ is diagonal in the $S^z$ representation. 
Thus the products $(S^z)^{2S+1-m}(S^+)^m$ are replaced by  
\be (S^z)^{2S+1-m}(S^+)^m=\sum_{i=0}^{2S-m}\delta_i^{(S,m)}\,
(S^z)^i\;\;(S^+)^m\,. \label{e13} \ee

\noindent 7) The corresponding reduction of spin operator products 
with an arbitrary order of the operators $S^+$, $S^-$, and $S^z$,
for example $(S^z)^{2S+1-m}(S^-)^m$, can be deduced
 from Eqs.(\ref{e12}),(\ref{e13}) by the usual commutation relations 
of the spin operators. Products consisting both $S^+$ and $S^-$ may be 
facilitated with the help of Eq.(\ref{e7}). If the total power 
is larger than $2S+1$, at first a particular subset of spin operators 
can be reduced. Then by repeated application of the procedure the 
complete spin operator product is replaced by a sum of products with 
a maximal total power equal to $2S$. 

\noindent 8) At the end of this derivation we like to remark a few 
important points. \\
(i) The described procedure for the reduction of the local spin 
operator products is \em exact, \em no approximation has been used. 
Products of spin operators located on different lattice sites, for 
example due to the exchange interaction, 
cannot be considered by the procedure outlined here. \\ 
(ii) The above results are valid for a sum of powers 
equal to or larger than $2S+1$.  The operators $(S^\pm)^m$ in 
Eqs.(\ref{e12}),(\ref{e13}) 
must not be cancelled, since this will lead to wrong results.  
For instance, cancellation of the common factor $(S^-)^m$ in 
Eq.(\ref{e12}) leads to the relation between powers of $S^z$ given by 
$(S^z)^{2S+1-m}=\sum_{i=0}^{2S-m}\delta_i^{(S,m)}\,(S^z)^i$. 
However, this relation is \em only \em correct for $m=0$, 
i.e.\ Eq.(\ref{e3}) !  \\
(iii) Facilitating the powers $(S^z)^{l+1}$ by replacing the products 
$S^-S^+$ on the right hand side of Eq.(\ref{e8}) -- (\ref{e10}) 
with the help of 
Eq.(\ref{e3}) does not help since this leads to identities. \\ 
(iv) Similar as for the coefficients $\alpha$ in Eq.(\ref{e3}) there 
should exist also recursion relations for the coefficients $\beta$, 
$\gamma$, and $\delta$  of Eqs.(\ref{e8}) -- (\ref{e13}), 
connecting respective coefficients for different spin 
quantum numbers $S$. We have not taken here the effort to deduce them. 

In conclusion, in this study we have shown how the total power of 
arbitrary 
products of spin operators $(S^+)^l(S^-)^m(S^z)^n$ can be reduced 
by one unit, if the sum of its powers is equal to $2S+1$. 
In addition we emphasize that by a repeated application of the
procedure \em all \em arbitrary spin operator products 
with a total power larger than $2S+1$ can be replaced by a sum 
of spin operators with a maximum total power of $2S$. 
These transformations are exact. It is important to note that the 
procedure outlined here is valid for \em local \em spin operator 
products only, 
i.e.\ all involved spin operators belong to the same lattice site. 
For example, the magnetic anisotropies, determining the direction of 
the magnetization, are usually single-ion quantities. 
We note that the consideration of such a single-ion anisotropy of
$l$-th order are only supported by spin quantum numbers $2S\ge l$
\cite{Mil95,BCS72,BMO86,CaC63}. Vice versa, for a given $S$ the
maximum order 
of distinct anisotropy terms as considered in the Heisenberg Hamiltonian 
is limited since anisotropy orders larger than $2S$ can be
replaced by lower order ones. This is due to exactly the same
circumstances leading to the reduction of spin operator products outlined
here. Note also that single-ion anisotropy terms can be replaced by
irreducible tensor operators \cite{Mil95,BCS72,BMO86,CaC63}. 
Furthermore, the investigation of the magnetic properties 
within a Green's function theory can be facilitated 
\cite{ErM91,Cal63,Tya67,FJK99}. 
For instance, the higher order Green's functions appearing in the
equation of motion can be replaced by lower order ones without
approximation, if the involved \em local \em spin operators are reduced
along the lines described here. 

Localized, quantum mechanical spins are present in the strongly 
correlated rare earth magnets, in which the localized $4f$- electrons 
cause the magnetic state. On the other hand, the $3d$- magnets are 
itinerant electron systems, the electrons causing the magnetism are 
not as localized as in the $4f$- systems, and the quantum mechanical 
nature of the \em equilibrium state \em of the magnetic properties 
may be hidden. This can be seen for example from the arbitrary values 
of the resulting magnetic moments. Consequently, its magnetic 
properties are usually represented by three-dimensional \em classical 
\em spin vectors $S\to\infty$. However, the quantum mechanical nature 
of the magnetism of these band magnets should be visible in the 
ultrashort time domain, for example within the magnetic relaxation 
after an excitation \cite{KBB99}. Such ultrashort temporal properties 
in the fsec- time domain can nowadays be investigated experimentally 
\cite{HMK97,ABP97}. For the theoretical study of these time dependent 
magnetic properties, which can be accomplished e.g.\ by application of
a \em time dependent \em Green's function (Keldysh-) formalism \cite{Kel65}, 
the reduction of the spin operator products may be a very useful 
tool. \\

\noindent Acknowledgements: P.J.J.\ gratefully acknowledges financial 
support from CONACyT, Grant No.\ 6-25851-E, and the hospitality of the 
Instituto de F\'\i sica de UASLP, Mexico. 
 
\newpage

\begin{sideways}
\begin{tabular}{|c|cccccc|ccccc|cccc|ccc|cc|c|} \hline
\ruo & \multicolumn{6}{c|}{$m=0$} & \multicolumn{5}{c|}{$m=1$} &
\multicolumn{4}{c|}{$m=2$} & \multicolumn{3}{c|}{$m=3$} 
& \multicolumn{2}{c|}{$m=4$} & $\;m=5$ \ruu \\ \hline
\ruo $\;S\;$ & 
$\;\delta_0^{(S,0)}$ & $\delta_1^{(S,0)}$ & $\delta_2^{(S,0)}$ & 
$\delta_3^{(S,0)}$ & $\delta_4^{(S,0)}$ & $\delta_5^{(S,0)}$ & 
$\;\delta_0^{(S,1)}$ & $\delta_1^{(S,1)}$ & $\delta_2^{(S,1)}$ & 
$\delta_3^{(S,1)}$ & $\delta_4^{(S,1)}$ & $\;\delta_0^{(S,2)}$ & 
$\delta_1^{(S,2)}$ & $\delta_2^{(S,2)}$ & $\delta_3^{(S,2)}$ & 
$\delta_0^{(S,3)}$ & $\delta_1^{(S,3)}$ & $\delta_2^{(S,3)}$ & 
$\;\delta_0^{(S,4)}$ & $\delta_1^{(S,4)}$ & $\;\delta_0^{(S,5)}$ 
\ruu \\ \hline 
\ruo $\frac{1}{2}$ & $\frac{1}{4}$ & $0$ & - & - & - & - &
$\frac{1}{2}$ & - & - & - & - & - & - & - & - & - 
& - & - & - & - & - \\ 
\ruo 1 & 0 & 1 & 0 & - & - & - & 0 & 1 & - & - & - 
& 1 & - & - & - & - & - & - & - & - & - \\ 
\ruo $\frac{3}{2}$ & -$\frac{9}{16}$ & 0 & $\frac{10}{4}$ & 0 & - & - &
-$\frac{3}{8}$ & $\frac{1}{4}$ & $\frac{3}{2}$ & - & - & 
-$\frac{3}{4}$ & 2 & - & - & $\frac{3}{2}$ & - & - & - & - & - \\
\ruo 2 & 0 & -4 & 0 & 5 & 0 & - & 0 & -2 & 1 & 2 & - 
& 0 & -2 & 3 & - & -2 & 3 & - & 2 & - & - \\
\ruo $\frac{5}{2}$ & $\frac{225}{64}$ & 0 & -$\frac{259}{16}$ & 0 & 
$\frac{35}{4}$ & 0 & $\frac{45}{32}$ & -$\frac{9}{16}$ & 
-$\frac{25}{4}$ & $\frac{5}{2}$ & $\frac{5}{2}$ &
$\frac{15}{16}$ & -1 & -$\frac{7}{2}$ & 4 &
$\frac{15}{8}$ & -$\frac{23}{4}$ & $\frac{9}{2}$ &
-$\frac{15}{4}$ & 4 & $\frac{5}{2}$ \ruu \\ \hline
\end{tabular} 
\end{sideways} \hspace{0.5cm}
\begin{sideways}
Table 1: Coefficients $\delta_i^{(S,m)}$ occurring in 
Eqs.(\ref{e12}),(\ref{e13}) 
for several values of the spin quantum number $S=1/2,1,3/2,2,5/2$. 
\end{sideways}


\newpage 

\begin{references}
\bibitem[*]{pjj} On leave from Hahn- Meitner- Institut, Glienicker 
Str.100, D-14 109 Berlin, Germany.
\bibitem{BlH94} J.A.C. Bland, B. Heinrich, \em Ultrathin Magnetic 
Structures, \em (Springer Verlag, Berlin, 1994).
\bibitem{MSK98} 
Y. Millev, R. Skomski, J. Kirschner, Phys. Rev. B 58 (1998) 6305.
\bibitem{LRP93} G. Lugert, W. Robl, L. Pfau, M. Brockmann,
G. Bayreuther, J. Magn. Magn. Mater. 121 (1993) 498. 
\bibitem{FMA97} M. Farle, B. Mirwald-Schulz, A.N. Anisimov, W. Platow,
K. Baberschke, Phys. Rev. B 55 (1997) 3708. 
\bibitem{HBW97} O. Hjortstam, K. Baberschke, J.M. Wills, B. Johansson,
O. Erickson, Phys. Rev. B 55 (1997) 15 026. 
\bibitem{UZW99} C. Uiberacker, J. Zabloudil, P. Weinberger, L. 
Szunyogh, C. Sommers, Phys. Rev. Lett. 82 (1999) 1289. 
\bibitem{Cal66} H.B. Callen, E.R. Callen, J. Phys. Chem. Solids 
27 (1966) 1271. 
\bibitem{JeB98} P.J. Jensen, K.H. Bennemann, 
{\it Magnetism and Electronic Correlations in Local-Moment Systems: 
Rare-Earth Elements and Compounds'}, p.\ 113 (M. Donath, 
P.A. Dowben, W. Nolting, eds.; World Scientific, Singapore, 1998). 
\bibitem{foot} We note that the measurements for the anisotropies may 
depend on the experimental setup, e.g.\ whether 
a strong external magnetic field is present or not. Furthermore, 
in two-dimensional magnetic systems such a for thin films or 
multilayers the anisotropies have a very important
effect because they induce a magnetized state with a long range 
ordering at finite temperatures. See, e.g., C. Herring, C. Kittel, 
Phys. Rev. 81 (1951) 869; S.V. Maleev, Sov. Phys. JETP 43 (1976) 1240; 
V.L. Pokrovsky, M.V. Feigel'man, Sov. Phys. JETP 45 (1977) 291. 
\bibitem{Mil95} Y. Millev, M. F\"ahnle, Phys. Rev. B 51 (1995) 2937.  
\bibitem{HuU97} A. Hucht, K.D. Usadel, Phys. Rev. B 55 (1997) 12 309.
\bibitem{BCS72} H.A. Buckmaster, R. Chatterjee, and Y.H. Shing, 
  Phys.Stat.Sol.(a) 13 (1972) 9. 
\bibitem{HoP40} T. Holstein, H. Primakoff, Phys. Rev. 58 (1940) 1098. 
\bibitem{ErM91} R.P. Erickson, D.L. Mills, Phys. Rev. B 43 (1991) 
10 715; \em ibid. \em 44 (1991) 11 825. 
\bibitem{Cal63} H.B. Callen, Phys. Rev. 130 (1963) 890. 
\bibitem{Tya67} S.V. Tyablikov, \em Methods in the quantum theory of
magnetism \em (Plenum Press, New York, 1967).  
\bibitem{FJK99} P. Fr\"obrich, P.J. Jensen, P.J. Kuntz,
Europ. J. Phys. B 13 (2000) 477. 
\bibitem{TGH98} C. Timm, S.M. Girvin, P. Henelius, A.W. Sandvik,
Phys. Rev. B 58 (1998) 1464.  
\bibitem{TiJ00} C. Timm, P.J. Jensen, submitted to Phys. Rev. B (2000). 
\bibitem{BMO86} G.J. Bowden, J.P.D. Martin, J. Oitmaa, 
J.Phys.C 19 (1986) 551. 
\bibitem{CaC63} E.R. Callen, H.B. Callen, 
Phys. Rev. 129 (1963) 578. 
\bibitem{KBB99} R. Knorren, K.H. Bennemann, R. Burgermeister, and
  M. Aeschlimann, submitted to Phys. Rev. B (1999). 
\bibitem{HMK97} J. Hohlfeld, E. Matthias, R. Knorren, 
K.H. Bennemann, Phys. Rev. Lett. 78 (1997) 4861. 
\bibitem{ABP97} 
M. Aeschlimann, M. Bauer, S. Pawlik, W. Weber, R. Burgermeister, 
D. Oberli, and H.C. Siegmann, Phys. Rev. Lett. 79 (1997) 5158. 
\bibitem{Kel65} L.V. Keldysh, Sov. Phys. JETP 20 (1965) 1018. 
\end{references}
\end{document}